\documentclass[runningheads]{llncs}
\usepackage{graphicx}
\usepackage{tikz,xcolor,hyperref,xurl}
\usepackage{comment}
\usepackage{subcaption}
\usepackage{tabularx}
\definecolor{lime}{HTML}{A6CE39}
\DeclareRobustCommand{\orcidicon}{%
	\begin{tikzpicture}
	\draw[lime, fill=lime] (0,0) 
	circle [radius=0.16] 
	node[white] {{\fontfamily{qag}\selectfont \tiny ID}};
	\draw[white, fill=white] (-0.0625,0.095) 
	circle [radius=0.007];
	\end{tikzpicture}
	\hspace{-2mm}
}

\foreach \x in {A, ..., Z}{%
	\expandafter\xdef\csname orcid\x\endcsname{\noexpand\href{https://orcid.org/\csname orcidauthor\x\endcsname}{\noexpand\orcidicon}}
}

\begin{document}
\title{Moral and emotional influences on attitude stability towards COVID-19 vaccines on social media\thanks{This work was supported in part by the Knight Foundation and the Office of Naval Research grant MURI: Persuasion, Identity, \& Morality in Social-Cyber Environments, N00014-21-12749. Additional support was provided by the Center for Computational Analysis of Social and Organizational Systems (CASOS) at Carnegie Mellon University. The views and conclusions contained in this document are those of the authors and should not be interpreted as representing the official policies, either expressed or implied, of the Knight Foundation,  Office of Naval Research, or the U.S. Government.}}
\titlerunning{Attitude stability on social media}
%
\author{Samantha C. Phillips\inst{1}\orcidA,
Lynnette Hui Xian Ng\inst{1} \orcidB,
Wenqi Zhou\inst{2}\orcidD, and
Kathleen M. Carley\inst{1} \orcidC}
\authorrunning{S.C. Phillips et al.}
%

\institute{CASOS Center, Software and Societal Systems Department \\
Carnegie Mellon University, 5000 Forbes Ave, Pittsburgh, PA 15213 \\
\email{samanthp, huixiann, carley\}@andrew.cmu.edu} \and
Palumbo-Donahue School of Business \\ Duquense University, 600 Forbes Ave, Pittsburgh, PA 15282 \\
\email{zhouw@duq.edu}}

\maketitle              
\begin{abstract}
Effective public health messaging benefits from understanding antecedents to unstable attitudes that are more likely to be influenced.
This work investigates the relationship between moral and emotional bases for attitudes towards COVID-19 vaccines and variance in stance. 
Evaluating nearly 1 million X users over a two month period, we find that emotional language in tweets about COVID-19 vaccines is largely associated with more variation in stance of the posting user, except anger and surprise.
The strength of COVID-19 vaccine attitudes associated with moral values varies across foundations.
Most notably, liberty is consistently used by users with no or less variation in stance, while fairness and sanctity are used by users with more variation.
Our work has implications for designing constructive pro-vaccine messaging and identifying receptive audiences.
\keywords{Attitude strength  \and Emotion \and Moral values \and Social media.}
\end{abstract}
\section{Introduction}
Moral values and emotions are ubiquitous experiences that shape how we process information and form judgments. 
Efforts to curb public opinion, such as promoting the acceptance of COVID-19 vaccines, can leverage moral and emotional reasoning to appeal to target audiences \cite{chou2020considering,nan2022public}.
This practice can be more effective when the moral values or emotions are relevant to members of the target audience, that is, anti-vaccine belief holders \cite{amin2017association,rossen2019accepters,schmidtke2022evaluating,sun2021impact}.
Furthermore, moral values and emotions associated with an attitude can affect the stability of and openness to updating it.
For example, attitudes with a moral basis have been associated with more attitude-consistent behavior and resistance to persuasion \cite{aramovich2012opposing,luttrell2016making}.
Similarly, emotional reactions to attitude-relevant information have been associated with greater attachment to that belief \cite{visser2016attitude}.
Understanding antecedents of attitude strength can reveal which individuals are most receptive to public health messaging.

Social media provides insight into how people’s public-facing views change over time, and the corresponding change in justification for those views.
In this work, we examine the relationship between the use of moral and emotional language in tweets about COVID-19 vaccines and the variance of stance towards COVID-19 vaccines expressed by the posting user.
The variance of stance across two months is used as a proxy for attitude strength, which is characterized by stability over time, resistance to persuasion, and influence on behavior and information processing \cite{krosnick2014attitude}.
Through an empirical analysis the tweets of nearly 1 million X (formerly Twitter) users who tweeted about COVID-19 vaccines at least twice in a two month period, we aim to address the following:
\begin{quote}
    \emph{RQ1}: How is moral and emotional language used differently in pro- and anti-COVID-19 vaccine tweets?\\
    \emph{RQ2}: How is moral and emotional language used differently by X users who indicate variation in expressed stance versus those who have no variation in expressed stance? \\
    \emph{RQ3}: Do X users who include more moral and emotional language in their tweets about COVID-19 vaccines have more or less variation in stance towards COVID-19 vaccines?
\end{quote}
The selected time period between March and May of 2021 coincides with some of the highest vaccination rates in the United States throughout the COVID-19 pandemic\footnote{\url{https://usafacts.org/visualizations/covid-vaccine-tracker-states/}}.
For each tweet, we applied a lexicon-based methodology to detect moral foundations and emotions.
In addition, we developed and applied a classifier to label the stance of each tweet towards COVID-19 vaccines as \{pro, anti, neutral\}.
We then calculated the average use of each moral value and emotion, as well as the standard deviation of stances, across tweets for each user.

In \emph{RQ1}, we provide an overview of which moral foundations and emotions are used in pro- and anti-COVID-19 vaccine tweets.
In \emph{RQ2}, we compare the moral and emotional language used by the 310,258 users who expressed a single stance about COVID-19 vaccines with the 603,158 users with any variation in stance represented in their tweets. 
In addition, in \emph{RQ3}, we identify associations between moral and emotional language and stance variance among users with any variation in stance, excluding users with zero variation.  By separating our analysis in this fashion, the outcome variable is approximately normally distributed in the regression model in \emph{RQ3} but we can still evaluate distinct moral and emotional language used by users with no variance in stance.

\section{Related works}
Extensive previous work has examined associations between moral values and judgments.
Moral foundation theory proposes six automatic intuitions (i.e., foundation or value) that affect judgments \cite{graham2011mapping}.  These foundations include: care (well-being of others), fairness (justice), sanctity (purity), liberty (freedoms), loyalty (in-group/out-group relations), and authority (following rules and traditions).
Moral foundations most associated with vaccine hesitancy are purity and liberty, followed by authority \cite{amin2017association}.
In addition, vaccine hesitancy is associated with less need for care \cite{schmidtke2022evaluating,amin2017association}.
However, these studies simply ask people about their automatic moral intuitions and attitudes towards vaccines, so there is no assessment of changes in stance over time or how moral reasoning is used to justify judgments about vaccines specifically.
Nonetheless, they demonstrate existing interest in how moral values affect attitudes towards vaccines.

Like moral intuitions, emotions can affect information processing and attitude formation.
While there are a multitude of ways to conceptualize emotions, we use Plutchik's six-emotion model: happy, sad, anger, fear, disgust, surprise \cite{plutchik2001nature}. 
This model has been used to study emotions on social media \cite{ng2021bot,liu2024emotion}.
Certain emotional reactions to vaccine-relevant events can predict vaccine hesitancy, like anger and anxiety \cite{sun2021impact}.
We build on this work to assess which emotions are used in pro- and anti-vaccine tweets.

Scholars have evaluated moral aspects of an attitude as an antecedent to attitude strength, finding attitudes with a moral basis are more resistant to influence \cite{luttrell2016making,aramovich2012opposing} and predict consistent behavior \cite{luttrell2016making}.
Furthermore, emotional responses to attitude-relevant information typically indicate a high degree of importance has been placed on that view, implying attitude strength \cite{visser2016attitude,miller2016impact}.
We investigate if using emotional language to discuss a stance is also associated with stability over time.
Unlike previous analyses of antecedents of attitude stability, we use stance variability in social media posts over time as a measure of attitude strength, extending these works to consider stance over time outside a laboratory environment.
Taking these studies together, we expect that moral and emotional language are associated with \textit{less} variation in stance towards COVID-19 vaccines.



\section{Data and Methods}

\subsection{Dataset}
Our dataset was initially collected using a streaming keyword search via Twitter v1 API between March 12, 2021 thru May 11, 2021. We collected tweets that contained at least one of the following terms: coronavirus, Wuhan virus, Wuhanvirus, 2019nCoV, NCoV, NCoV2019, covid-19, covid19, covid 19.  We then further filtered the tweets to select those about vaccines using the following keywords: vaccine, vax, mrna, autoimmuneencephalitis, vaccination, getvaccinated, covidisjustacold, autism, covidshotcount, dose1, dose2, VAERS, GBS, believemothers, mybodymychoice, thisisourshot, killthevirus, proscience, immunization, gotmyshot, igottheshot, covidvaccinated, beatcovid19, moderna, astrazeneca, pfizer, johnson \& johnson, j\&j, johnson and johnson, jandj.  

In total, we extracted 2,283,281 users, 1,369,865 of which only tweeted one time about COVID-19 vaccines during this time period.
In this work, we analyze the remaining 913,416 users who tweeted at least twice in our time period for a total of 6,811,854 tweets.


\subsection{Stance detection}
To evaluate stance consistency, we first developed a classifier to label the stance (pro, anti, neutral) of each tweet towards COVID-19 vaccines. 
To generate a training dataset, two annotators labelled 1034 tweets with Cohen's Kappa of 0.741 and agreement of 84.2\%.  A third annotator labelled the 163 tweets the initial two annotators disagreed on.  
We removed the 6 tweets that all three disagreed on. 
We then fine-tuned a BERTweet model\footnote{\url{https://huggingface.co/docs/transformers/model\_doc/bertweet}} for the stance detection task to obtain an accuracy of 72.8\% on the evaluation data. The stance classifier returns a stance label and confidence score for each tweet.

We applied this stance detection classifier to original tweets, retweets, replies, and quote tweets in our dataset.
This gives us 3,547,645 pro-vaccine, 2,928,820 neutral, and 335,389 anti-vaccine tweets.
To obtain a measure of variation in stance per user, we calculated the standard deviation of stance expressed across their tweets.  
Of the 913,416 users in our dataset, 310,258 express a consistent stance.
Removing the 310,258 users with a standard deviation of 0, the average standard deviation is 0.62 (std. dev. 0.18) and median is 0.58.

\subsection{Moral value and emotion extraction}

We use the Netmapper software to extract words and phrases associated with each moral value (care, fairness, liberty, sanctity, loyalty, authority) and emotion (happiness, sadness, fear, anger, disgust, surprise) \cite{carley2018ora,tausczik2010psychological}. 
We recorded the number of concepts representing each moral/emotion variable in each tweet.

\section{Results}

\subsection{RQ1: Moral values and emotions in pro- and anti-COVID-19 vaccine tweets}
\begin{figure}
    \centering
    \includegraphics[width=1\linewidth]{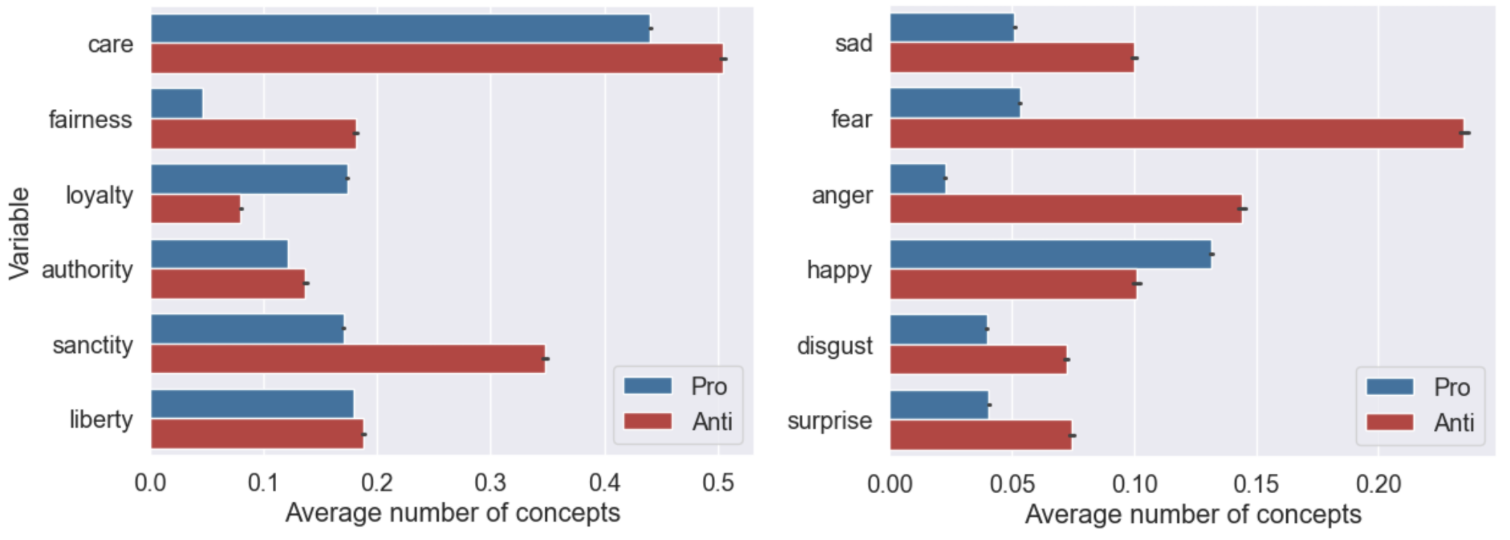}
    \caption{Average and 95\% CI number of concepts associated with each moral foundation (left) and emotion (right) in pro- and anti-COVID-19 vaccine tweets.}
    \label{fig:fig1}
\end{figure}

Figure \ref{fig:fig1} displays the average and 95\% confidence interval of the number of concepts associated with each moral value and emotion in pro- and anti-COVID-19 vaccine tweets.
Overall, anti-vaccine tweets contain more moral and emotional language than pro-vaccine tweets for most types of moral values and emotions.  
That is, anti-vaccine tweets contain more references to care, fairness, authority, sanctity, and liberty foundations than pro-vaccine tweets.
On the other hand, pro-vaccine tweets contain more loyalty terms.
Anti-vaccine tweets are also more likely to contain sad, fear, anger, disgust and surprise emotions, while pro-vaccine tweets contain more happiness.
Table \ref{tab:ex} contains examples of tweets containing select moral values and stances.

\begin{table}[h]
    \centering
    \begin{tabularx}{\textwidth}{p{1.5cm}|p{3cm}|p{7.2cm}}
    \hline 
    Stance & Moral foundation    &  Example tweet \\ \hline
    pro     &  care & If you can't see the urgency of saving humanity by granting a \#TRIPSWaiver to ensure \#VaccineForAll , I don't see what else you see. Time to stop putting Health in the Marketplace.  \\ \hline
    anti & care & ``Vaccinating'' babies now with the mRNA jabs? How can this even be ethical?  \\ \hline
    pro & loyalty & His son has taken both doses of the COVID-19 vaccine, and hopefully was wearing a mask. They are both putting action behind their push towards herd immunity. I hope you’ll get the vaccine too! \\ \hline
    anti & sanctity & I am not taking the vaccine.  I take 5000 units of vitamin D3.  If God wants me to die from COVID-19 no vaccine will stop God's Will. \\ \hline
    \end{tabularx}
    \caption{Example tweets}
    \label{tab:ex}
\end{table}

\subsection{RQ2: Moral values and emotions in tweets by users expressing no versus any variation in stance}
Table \ref{tab:ttest} contains t-statistics from t-tests comparing the use of moral and emotional language by users with no versus any variance in expressed stance.
Users that are completely consistent in their expressed stance tend to use anger, loyalty, care, and liberty in their tweets more than users that express variation in their stance.  
Users who express any variation in stance are more likely to use happy, disgust, surprise, fear, or sad emotional language.  Furthermore, they tend to refer to sanctity, authority and fairness foundations more in their tweets.

\begin{table}[h]
    \centering
    \begin{tabular}{lc|cc}
    \hline
    Emotion & & Moral value &  \\
    \hline
        Variable &  T-statistic & Variable &  T-statistic  \\
        \hline
       Anger  & -44.22   & Loyalty & -37.95   \\
    Happy & 6.36  & Care & -34.36   \\
    Disgust & 17.26  & Liberty & -22.57   \\
 Surprise & 30.18 & Sanctity & 21.90    \\
    Fear & 32.3 & Authority & 27.42  \\
    Sad & 37.86  & Fairness & 44.43   \\
    \hline
    \end{tabular}
    \caption{T-statistics from t-tests comparing the average number of concepts associated with each moral value and emotion in tweets by users who expressed no variation versus any variation in stance towards COVID-19 vaccines.  Negative values indicate the variable is used more in tweets by users that express no variation in stance.  All t-tests are significant ($p<0.0001$).}
    \label{tab:ttest}
\end{table}

\subsection{RQ3: Moral values and emotions in tweets by users expressing some degree of variation in stance}
\begin{figure}
    \centering
    \includegraphics[width=1\linewidth]{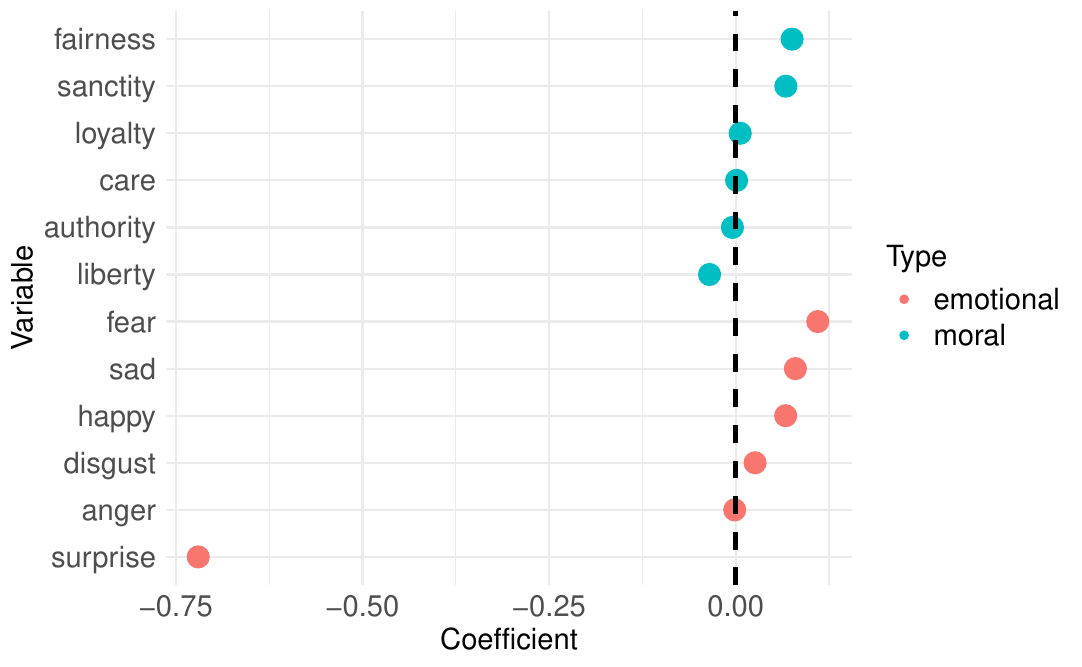}
    \caption{Regression coefficients for the standard deviation of expressed stance across tweets.  Negative values indicate the variable is associated with less variation in stance.  Error bars indicate the 95\% confidence intervals.  All coefficients are significant ($p<0.0001$) except for anger ($p>0.1$) and care ($p=0.049$).}
    \label{fig:regression}
\end{figure}

Figure \ref{fig:regression} displays the coefficients for an OLS regression model predicting the standard deviation of expressed stance across tweets for each user given the use of moral and emotional language in their tweets about COVID-19 vaccines.  All coefficients are significant ($p<0.0001$) except for anger ($p>0.1$).

More emotional language is largely associated with more variation in stance, except surprise.  
Liberty and loyalty are associated with less variation in stance, while the remaining moral foundations (fairness, sanctity, authority, care) are associated with more variation.

\section{Discussion}
The aim of this work is to investigate how moral foundations and emotions associated with an attitude affect the stability of that attitude in a social media context.
To do so, we analyzed tweets expressing a stance toward COVID-19 vaccines.

First, we examined which moral foundations and emotions are used in tweets expressing pro- and anti-COVID-19 vaccine views to provide an overview of the context in which each feature is typically used.
The findings in \emph{RQ1} indicate that anti-vaccine tweets tend to contain more moral and emotional language than pro-vaccine tweets except for loyalty and happiness.
Many pro-vaccine tweets that contain language associated with loyalty contain information about where a community can access vaccines or celebrate that vaccines were successfully given to a community (e.g., Americans, indigenous tribes in Oklahoma).
As one may expect, vaccine hesitant messaging tends to use more negative emotional language than those expressing support. 

Our findings echo previous work showing vaccine hesitancy is associated with moral values like sanctity and liberty \cite{amin2017association}.
Anti-vaccine tweets that use language associated with sanctity include concerns about the efficacy and safety of COVID-19 vaccines \cite{amin2017association}, and use more negative emotional language \cite{faasse2016comparison}.
However, while  vaccine hesitancy has been associated with lack of need for care \cite{schmidtke2022evaluating,amin2017association}, we find care is the most used moral foundation in anti-vaccine tweets.
These tweets include concern about how negative side effects from vaccines are affecting or may affect others.

In addressing \emph{RQ2} and \emph{RQ3}, we find emotional language is associated with more variation in stance, except for anger and surprise.  
That is, certain emotions (fear, sadness, happiness, disgust) indicate unstable attitudes.
Users may be less tied to their view once their emotional reaction is less salient.
Indeed, studies show that both positive and negative emotions can override analytical thinking processes, increasing the effect of biased information processing on truth discernment \cite{scheufele2019science,sinclair2010affective}.

Conversely, anger is associated with consistent stance, but the association between anger and stance variance among those who do vary to some degree is not significant.  
This highlights the different ways different emotions can impact judgments depending on the individual and context.
Surprise is most associated with users that express some, but small, variance in stance.
Surprise drives belief updating by attracting attention to unexpected information, motivating people to reconsider and potentially rectify their understanding \cite{maguire2011making}.

The role of moral language varies across stance foundations.  Liberty is consistently used by users with no or less variation in stance, while fairness and sanctity are used by users with more variation in stance towards COVID-19 vaccines.
Care and loyalty are associated with no or high variation, while authority is associated with some but low variation. 
The same moral value can play a very different role depending on the context and individual differences. 
For example, moral identity is how important the individual considers being a ``moral'' person to their self-concept. 
Prioritizing ``binding'' values (loyalty, authority, sanctity) represent a group-oriented view of morality.  
Adherence to binding foundations is associated with out-group derogation if moral identity is weak and willingness to help out-group members if moral identity is strong \cite{smith2014moral}.
More work is needed to identify variables influencing the effect of moral values on stance consistency.


The results of this work inform how moral values and emotions associated with a belief affect the stability of that belief. 
Our work has implications for constructing effective campaigns to promote vaccines and debunk vaccine misinformation, as well as directing these messages to receptive audiences.
Specifically, promoting and tailoring pro-vaccine messaging to people expressing moral and emotional language associated with less attitude stability (i.e., sanctity, fairness, fear, sadness, happiness, disgust) can help to optimize public health outreach.

\subsection{Future work and limitations}
Our study is limited to the platform X.
Future work should assess the use of moral values and emotions towards vaccine attitudes on other social media platforms. 
These analyses could provide vital insight into how organizations designing public health messaging should anticipate shifts in motivations as vaccines are developed and approved for the public.  
Furthermore, we did not attempt to identify automated accounts (``bots'') in our analysis \cite{ng2021bot}.
Future work should examine the extent to which removing users displaying bot-like behavior affects the observed relationship between moral-emotional bases for an attitude and the consistency of that attitude.

Rather than separately examining which moral values and emotions predict more or less vaccine hesitancy over time, we looked at any change in stance regardless of the direction.
Disentangling the effects on increases and decreases in anti-vaccine attitudes can directly inform public health messaging aiming to increase vaccinations \cite{schmidtke2022evaluating}.

Lastly, attitudes towards vaccines are multi-faceted, complex judgments that often include far more nuance that we include in this work.
Indeed, vaccine hesitancy can be motivated by a wide range of factors, such as concerns over vaccine testing protocols, institutional distrust, and belief in alternatives \cite{weinzierl2022hesitancy}.
Considering more fine-grained stance categories may reveal important distinctions in the role of moral values and emotions in stance (in)consistency.

\section{Conclusion}
We studied the influence of moral and emotional associations with an attitude on the stability of that attitude over time.
Specifically, we examined the moral values, emotions, and stance expressed in tweets by nearly one million users on X  over a two month period.
We find that (1) moral and emotional language is used more in anti-vaccine tweets, except for happiness and loyalty. 
(2) Emotional language is associated with larger variation with stance, except for anger and surprise, which are associated with consistent stance. 
(3) The role of moral language and stance variability differs across values, e.g. liberty is associated with consistent stances while fairness and sanctity are associated with larger stance variation.
Our work shows that moral and emotional reasoning in attitudes can be used to predict receptive audiences, which can be especially useful in counter-misinformation campaigns.

\bibliographystyle{splncs04}
\bibliography{biblio}

\begin{thebibliography}{10}
\providecommand{\url}[1]{\texttt{#1}}
\providecommand{\urlprefix}{URL }
\providecommand{\doi}[1]{https://doi.org/#1}

\bibitem{amin2017association}
Amin, A.B., Bednarczyk, R.A., Ray, C.E., Melchiori, K.J., Graham, J., Huntsinger, J.R., Omer, S.B.: Association of moral values with vaccine hesitancy. Nature Human Behaviour  \textbf{1}(12),  873--880 (2017)

\bibitem{aramovich2012opposing}
Aramovich, N.P., Lytle, B.L., Skitka, L.J.: Opposing torture: Moral conviction and resistance to majority influence. Social Influence  \textbf{7}(1),  21--34 (2012)

\bibitem{carley2018ora}
Carley, L.R., Reminga, J., Carley, K.M.: Ora \& netmapper. In: International conference on social computing, behavioral-cultural modeling and prediction and behavior representation in modeling and simulation. Springer. vol.~3, p.~7 (2018)

\bibitem{chou2020considering}
Chou, W.Y.S., Budenz, A.: Considering emotion in covid-19 vaccine communication: addressing vaccine hesitancy and fostering vaccine confidence. Health communication  \textbf{35}(14),  1718--1722 (2020)

\bibitem{faasse2016comparison}
Faasse, K., Chatman, C.J., Martin, L.R.: A comparison of language use in pro-and anti-vaccination comments in response to a high profile facebook post. Vaccine  \textbf{34}(47),  5808--5814 (2016)

\bibitem{graham2011mapping}
Graham, J., Nosek, B.A., Haidt, J., Iyer, R., Koleva, S., Ditto, P.H.: Mapping the moral domain. Journal of personality and social psychology  \textbf{101}(2), ~366 (2011)

\bibitem{krosnick2014attitude}
Krosnick, J.A., Petty, R.E.: Attitude strength: An overview. Attitude strength pp. 1--24 (2014)

\bibitem{liu2024emotion}
Liu, Z., Zhang, T., Yang, K., Thompson, P., Yu, Z., Ananiadou, S.: Emotion detection for misinformation: A review. Information Fusion p. 102300 (2024)

\bibitem{luttrell2016making}
Luttrell, A., Petty, R.E., Bri{\~n}ol, P., Wagner, B.C.: Making it moral: Merely labeling an attitude as moral increases its strength. Journal of Experimental Social Psychology  \textbf{65},  82--93 (2016)

\bibitem{maguire2011making}
Maguire, R., Maguire, P., Keane, M.T.: Making sense of surprise: an investigation of the factors influencing surprise judgments. Journal of Experimental Psychology: Learning, Memory, and Cognition  \textbf{37}(1), ~176 (2011)

\bibitem{miller2016impact}
Miller, J.M., Krosnick, J.A., Holbrook, A., Tahk, A., Dionne, L.: The impact of policy change threat on financial contributions to interest groups. Political psychology: New explorations pp. 172--202 (2016)

\bibitem{nan2022public}
Nan, X., Iles, I.A., Yang, B., Ma, Z.: Public health messaging during the covid-19 pandemic and beyond: Lessons from communication science. Health communication  \textbf{37}(1),  1--19 (2022)

\bibitem{ng2021bot}
Ng, L.H.X., Carley, K.M.: Bot-based emotion behavior differences in images during kashmir black day event. In: International Conference on Social Computing, Behavioral-Cultural Modeling and Prediction and Behavior Representation in Modeling and Simulation. pp. 184--194. Springer (2021)

\bibitem{plutchik2001nature}
Plutchik, R.: The nature of emotions: Human emotions have deep evolutionary roots, a fact that may explain their complexity and provide tools for clinical practice. American scientist  \textbf{89}(4),  344--350 (2001)

\bibitem{rossen2019accepters}
Rossen, I., Hurlstone, M.J., Dunlop, P.D., Lawrence, C.: Accepters, fence sitters, or rejecters: Moral profiles of vaccination attitudes. Social Science \& Medicine  \textbf{224},  23--27 (2019)

\bibitem{scheufele2019science}
Scheufele, D.A., Krause, N.M.: Science audiences, misinformation, and fake news. Proceedings of the National Academy of Sciences  \textbf{116}(16),  7662--7669 (2019)

\bibitem{schmidtke2022evaluating}
Schmidtke, K.A., Kudrna, L., Noufaily, A., Stallard, N., Skrybant, M., Russell, S., Clarke, A.: Evaluating the relationship between moral values and vaccine hesitancy in great britain during the covid-19 pandemic: A cross-sectional survey. Social Science \& Medicine  \textbf{308},  115218 (2022)

\bibitem{sinclair2010affective}
Sinclair, M., Ashkanasy, N.M., Chattopadhyay, P.: Affective antecedents of intuitive decision making. Journal of Management \& Organization  \textbf{16}(3),  382--398 (2010)

\bibitem{smith2014moral}
Smith, I.H., Aquino, K., Koleva, S., Graham, J.: The moral ties that bind... even to out-groups: The interactive effect of moral identity and the binding moral foundations. Psychological science  \textbf{25}(8),  1554--1562 (2014)

\bibitem{sun2021impact}
Sun, R., Wang, X., Lin, L., Zhang, N., Li, L., Zhou, X.: The impact of negative emotional reactions on parental vaccine hesitancy after the 2018 vaccine event in china: a cross-sectional survey. Human Vaccines \& Immunotherapeutics  \textbf{17}(9),  3042--3051 (2021)

\bibitem{tausczik2010psychological}
Tausczik, Y.R., Pennebaker, J.W.: The psychological meaning of words: Liwc and computerized text analysis methods. Journal of language and social psychology  \textbf{29}(1),  24--54 (2010)

\bibitem{visser2016attitude}
Visser, P.S., Krosnick, J.A., Norris, C.J.: Attitude importance and attitude-relevant knowledge: Motivator and enabler. In: Political psychology, pp. 217--259. Psychology Press (2016)

\bibitem{weinzierl2022hesitancy}
Weinzierl, M.A., Harabagiu, S.M.: From hesitancy framings to vaccine hesitancy profiles: A journey of stance, ontological commitments and moral foundations. In: Proceedings of the International AAAI Conference on Web and Social Media. vol.~16, pp. 1087--1097 (2022)

\end{thebibliography}

\end{document}